\documentclass[preprint]{iucr}

\usepackage{amsmath, amsthm, amssymb, amsfonts, textcomp, gensymb, subfigure, import, color, siunitx, float, bm, graphicx, url}

     \journalcode{S}              

\begin{document}                  



\title{Mapping data between sample and detector conjugated spaces in Bragg coherent diffraction imaging}


\author[a]{David}{Yang}
\author[a]{Nicholas}{W. Phillips}
\cauthor[a]{Felix}{Hofmann}{david.yang@eng.ox.ac.uk  \linebreak Email: felix.hofmann@eng.ox.ac.uk}

\aff[a]{Department of Engineering Science, University of Oxford, Parks Road, Oxford, OX1~3PJ, \country{UK}}


\keyword{Bragg Coherent X-ray Diffraction Imaging, Crystal Reflection, Detector Conjugated Space, Simulation, Space Transformation}

\maketitle                        


\begin{abstract}
Bragg coherent X-ray diffraction imaging (BCDI) is a non-destructive, lensless method for 3D-resolved, nanoscale strain imaging in micro-crystals. A challenge, particularly for new users of the technique, is accurate mapping of experimental data, collected in the detector reciprocal space coordinate frame, to more convenient orthogonal coordinates, e.g. attached to the sample. This is particularly the case since different coordinate conventions are used at every BCDI beamline. The reconstruction algorithms and mapping scripts composed for individual beamlines are not readily interchangeable. To overcome this, a BCDI experiment simulation with a plugin script that converts all beamline angles to a universal, right-handed coordinate frame is introduced, making it possible to condense any beamline geometry into three rotation matrices. The simulation translates a user-specified 3D complex object to different BCDI-related coordinate frames. It also allows the generation of synthetic coherent diffraction data that can be inserted into any BCDI reconstruction algorithm to reconstruct the original user-specified object. Scripts are provided to map from sample space to detector conjugated space, detector conjugated space to sample space and detector conjugated space to detector conjugated space for a different reflection. This provides the reader with the basis for a flexible simulation tool kit that is easily adapted to different geometries. It is anticipated that this will find use in the generation of tailor-made supports for phasing of challenging data and exploration of novel geometries or data collection modalities.
\end{abstract}


\section{Introduction} \label{section:Introduction}
Bragg coherent X-ray diffraction imaging (BCDI) is an emerging technique in the field of lensless imaging, able to provide a 3D spatial resolution of less than 10 nm \cite{Cherukara2018}. Micro-crystal samples for BCDI are often produced by deposition of a thin film, followed by subsequent dewetting, such as Au \cite{Robinson2001}, or Pb \cite{Pfeifer2006}. Immobilization of micro-crystals on a substrate has also been employed \cite{Pfeifer2006,Harder2013,Hruszkewycz2017}, and more recently a top-down focused ion beam fabrication method has been demonstrated to allow manufacture of strain microscopy samples from any crystalline material \cite{Hofmann2019}. BCDI has been used to study strain in samples subject to a wide variety of increasingly complex conditions. For example, it has been used to investigate dislocations in crystal growth and dissolution cycles \cite{Clark2015}, strain in battery cathode nanoparticles \cite{Ulvestad2015} and acoustic phonons in zinc oxide crystals \cite{Ulvestad2017}.

To perform BCDI, a crystalline sample is illuminated by a spatially coherent synchrotron X-ray beam with incoming wavevector, $\mathbf{S}_\mathrm{0 lab}$. he sample must be sufficiently small to fit inside the coherent volume of $\mathbf{S}_\mathrm{0 lab}$, which limits the sample size to 1$\mathrm{\mu m}\times$1$\mathrm{\mu m}\times$1$\mathrm{\mu m}$ for BCDI \cite{Clark2012,Hofmann2017a}.  If the Bragg condition is met, the X-rays are scattered from the sample volume and interfere, resulting in a diffraction pattern or the Fourier transform, $\mathcal{F}$, of the complex electron density of the crystal in the far-field Fraunhofer diffraction regime \cite{Xiong2014}. Approximating atoms as points, a perfect, infinite crystal will produce a diffraction pattern resembling a lattice of Dirac delta functions, a perfect, finite crystal will produce a similar array of peaks but with a defined symmetric width, and an imperfect crystal will produce the previous pattern, but with asymmetric width \cite{Xiong2014}. These diffraction patterns are collected on a pixelated area detector positioned perpendicular to the outgoing wavevector, $\mathbf{S}_\mathrm{lab}$, in the far-field, intersecting part of the Ewald sphere for a specific hkl reflection at the Bragg condition \cite{Xiong2014}. By rotating the sample about a rocking axis, a series of diffraction patterns is collected as the reciprocal lattice point moves through the Ewald sphere. This represents slices in the $\mathbf{q'_{3}}$ direction (Fig. \ref{fig:geometry}) through the chosen 3D Bragg peak \cite{Robinson2001}. 

These diffraction pattern intensities do not contain any phase information, known as the phase loss problem \cite{Miao1999}. What we record is the square of the amplitude of the $\mathcal{F}$ of the complex electron density of the crystal \cite{Ulvestad2015,Robinson2001}.  By oversampling the diffraction pattern by at least twice the Nyquist frequency \cite{Miao2000} (at least 4 pixels per fringe), and applying geometric constraints, the phase can be recovered using iterative phase-retrieval algorithms \cite{Fienup1982}. Only after retrieving the phase can the 3D image be reconstructed by inverting the diffraction pattern through an inverse Fourier transform \cite{Miao2000,Robinson2001,Clark2012}. The resulting amplitude, $\mathbf{\rho(r)}$, and phase, $\mathbf{\psi(r)}$, where $\mathbf{r}$ is the position vector, can be interpreted as the complex electron density of the crystalline volume associated with the particular crystal reflection. $\mathbf{\psi(r)}$ corresponds to the projection of the lattice displacement field, $\mathbf{u(r)}$, onto the scattering vector, $\mathbf{Q}_\mathrm{lab}$, of the $hkl$ crystal reflection under consideration, i.e. $\mathbf{\psi(r)} = \mathbf{Q}_\mathrm{lab}\cdot\mathbf{u(r)}$  \cite{Robinson2009}.As such, BCDI provides information about both the shape of the scattering crystal domain as well as the lattice displacement field within it. If three or more linearly independent reflections are measured for a single crystal, multi-reflection BCDI (MBCDI) can be implemented to calculate the full lattice strain tensor, $\boldsymbol{\varepsilon} \mathbf{(r)}$, for the crystal by differentiating the full $\mathbf{u(r)}$ \cite{Newton2010,Hofmann2017b}.

BCDI is an increasingly popular technique, but its disadvantage is the relatively high barrier to entry. It requires access to coherent diffraction beamlines at select synchrotrons, each with different geometry conventions. Here we introduce a BCDI simulation program with a beamline-specific plugin script that simplifies the geometry for a beamline by converting all motor rotation angles to a right-handed convention and generates three fundamental matrices that fully describe the measurement (see Section \ref{section:Simulation_Geometry}). The simulation provides both the tools and the underlying theory required to move with ease between BCDI-related spaces. A challenge in the community is mapping between these spaces, particularly between detector conjugated space (DCS), where the measurement and phase retrieval are performed, and an orthogonal sample space (SS) attached to the object of study. We introduce three scripts that allow the reader to map from sample space to detector conjugated space, detector conjugated space to sample space, and detector conjugated space to detector conjugated space for a different crystal reflection. These scripts also incorporate the beamline-specific plugin, making them readily adaptable to any geometry.

\section{Simulation Geometry} \label{section:Simulation_Geometry}
\begin{figure} \label{fig:geometry}
    \centering
    \includegraphics[width=\textwidth,scale=0.5]{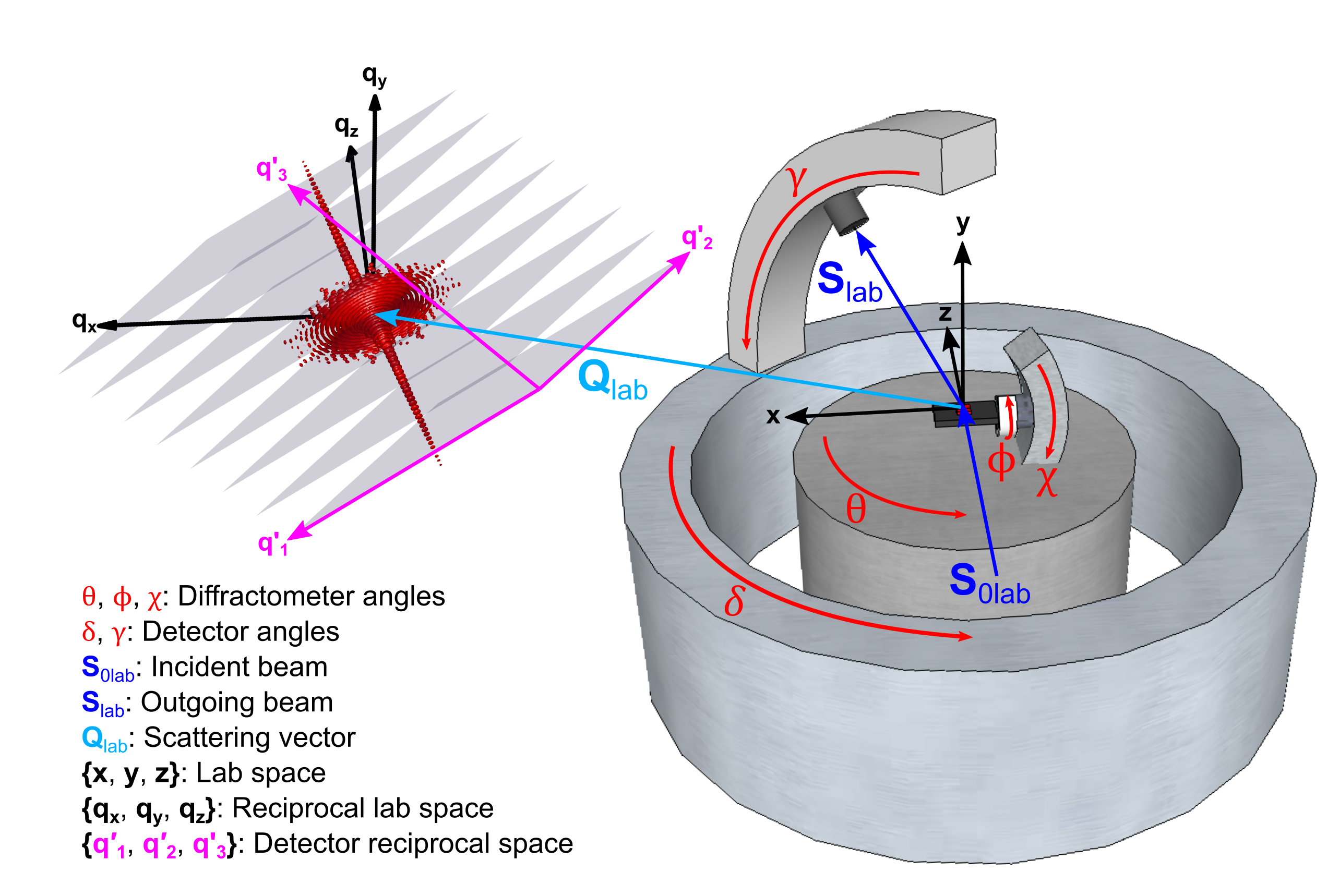}
    \caption{A schematic of the simulated experimental geometry showing 2D slices through the 3D Bragg peak as a consequence of rocking about the $y$-axis. Diffractometer angles, $\theta$, $\phi$, and $\chi$, as well as detector angles $\delta$ and $\gamma$, are labelled with the red arrows pointing in the positive direction of rotation. $\mathbf{S}_\mathrm{0 lab}$ and $\mathbf{S}_\mathrm{lab}$ represent the incident and reflected X-ray beams respectively. $\mathbf{Q}_\mathrm{lab}$ is the scattering vector. Three coordinate frames are shown: lab space ($\mathbf{x}$, $\mathbf{y}$ and $\mathbf{z}$), reciprocal space ($\mathbf{q}_\mathrm{x}$, $\mathbf{q}_\mathrm{y}$ and $\mathbf{q}_\mathrm{z}$) and detector reciprocal space ($\mathbf{q}'_\mathrm{1}$, $\mathbf{q}'_\mathrm{2}$ and $\mathbf{q}'_\mathrm{3}$).}
\end{figure}

This section covers the geometry that is used in the computer scripts and in Section \ref{section:Computer_Codes}. A schematic of the simulated geometry is shown in Fig. \ref{fig:geometry}. The laboratory coordinates are oriented such that the incident X-ray beam is in the direction of the positive $z$-axis, the $y$-axis is vertically up and the $x$-axis is outboard to the left. Here the following coordinates are used to describe the various spaces:

\begin{equation} \label{eq:spaces_axes} 
    \begin{aligned}
        \mathrm{Sample\ Space\ (SS)}:&\hspace{1mm} \mathbf{x}_\mathrm{sam},\hspace{1mm}\mathbf{y}_\mathrm{sam},\hspace{1mm}\mathbf{z}_\mathrm{sam},\\
        \mathrm{Lab\ Space\ (LS)}:&\hspace{1mm} \mathbf{x},\hspace{1mm}\mathbf{y},\hspace{1mm}\mathbf{z},\\
        \mathrm{Reciprocal\ Lab\ Space\ (RLS)}:&\hspace{1mm} \mathbf{q}_\mathrm{x},\hspace{1mm}\mathbf{q}_\mathrm{y},\hspace{1mm}\mathbf{q}_\mathrm{z},\\
        \mathrm{Detector\ Reciprocal\ Space\ (DRS)}:&\hspace{1mm} \mathbf{q}'_\mathrm{1},\hspace{1mm}\mathbf{q}'_\mathrm{2},\hspace{1mm}\mathbf{q}'_\mathrm{3},\\
        \mathrm{Detector\ Conjugated\ Space\ (DCS)}:&\hspace{1mm} \mathbf{x}',\hspace{1mm}\mathbf{y}',\hspace{1mm}\mathbf{z}'.
    \end{aligned}
\end{equation}
\\
The detector has two angles of rotation: $\delta$ and $\gamma$, which rotate about the $y$-axis and $x$-axis, respectively, when angles are set to zero as shown by \cite{Hofmann2017a}, and all rotations adhere to a right-handed convention. The diffraction pattern is generated as if the detector were looking down onto the sample (as opposed to a projection onto a film). Explicitly, this means that if there is a diffraction pattern on the detector screen, decreasing $\gamma$ will move the detector up in the $y$ or $\mathbf{q'_{2}}$ direction, and the peak will move down on the screen. Similarly, decreasing $\delta$ will move the detector to the right as $\mathbf{q'_{1}}$ increases and the peak will move left on the screen. The sample stage has three angles of rotation: $\chi$, $\phi$, and $\theta$, which rotate about the $z$-axis, $x$-axis, and $y$-axis respectively, when angles are set to zero, also shown by \cite{Hofmann2017a}. If the sample is rocked about the $y$-axis by an increment of $\Delta \theta$ from negative to positive, this corresponds to collecting slices through the Bragg peak along the $\mathbf{q'_{3}}$ direction. In the case of the instrumentation at 34-ID-C, angular limits imposed by stage travel ranges need to be considered, as certain geometries are inaccessible (Appendix \ref{appendix:a1}).

\sloppy Overall, the geometry of a BCDI experiment requires three rotation matrices to fully describe the affine transformation between coordinate systems: ${R_\mathrm{x,y,z}}$, ${R_{\Delta \mathrm{q'_{1,2}}}}$, and ${R_{\Delta \mathrm{q'_{3}}}}$, which correspond to the rotation matrices of the sample, detector and rocking axis respectively. Once defined, the BCDI simulation and mapping can be carried out for any beamline.

\section{Computer Codes} \label{section:Computer_Codes}
\sloppy A program, \texttt{BCDI\_{Simulation}}, found at \url{https://doi.org/10.5281/zenodo.3347113}, and three scripts, \texttt{SS\_to\_DCS}, \texttt{DCS\_to\_DCS} and \texttt{DCS\_to\_SS}, available at \url{https://doi.org/10.5281/zenodo.3347109}, are published with this work. For the most updated version of the codes, the reader is asked to \url{https://github.com/Hofmann-Group}. The recommended machine requirements for these codes are: MATLAB version 2018b or later, 8GB RAM, an Intel or AMD x86-64 processor with four cores, and at least 2GB of free hard disk space. The use of an advanced research computing facility will greatly reduce computation time. For instance the authors used the University of Oxford Advanced Research Computing (ARC) facility for this work \cite{Richards2015}. The maximum recommended complex double array size for these settings is 256$\times$256 $\times$256, which can be saved as a separate \texttt{.mat} file. In this paper, a script refers to a single MATLAB file, a program refers to a script that calls upon other scripts, and a code can refer to both scripts and programs.

    \subsection{Bragg Coherent Diffraction Imaging Simulation} \label{subsection:BCDI_Simulation}
    The program, \texttt{BCDI\_{Simulation}}, allows the simulation of a BCDI experiment starting with a user-defined shape (a 3D complex double array). Similar to a laboratory experiment, the reader is required to input the X-ray wavelength, pixel size, detector size, $hkl$ reflection (or detector and sample position angles), rocking increment, rocking axis and detector distance. Alternatively, the $\mathbf{UB}$ matrix can be input if micro-beam Laue diffraction was used to pre-determine the lattice orientation of the sample \cite{Hofmann2017a}. Using the inputs, the program translates and plots the shape in five different spaces: sample space (SS), lab space (LS), reciprocal lab space (RLS), detector reciprocal space (DRS) and detector conjugated space (DCS); these spaces are described in detail in Section \ref{section:Space_Transformations} and are shown in Fig. \ref{fig:spaces}. The simulation also produces a TIFF file containing 2D diffraction pattern slices through the simulated Bragg peak, which can be fed into a reconstruction algorithm, e.g. \cite{Clark2012} to reconstruct the user-defined shape in LS and DCS. Accordingly, the simulated shapes can be compared with the reconstructed shapes. At the time of writing, the code's geometry is based on beamline 34-ID-C at the Advanced Photon Source (APS) in Argonne National Laboratory, USA, because it is a heavily used instrument in the history of BCDI. This beamline uses both left- and right-handed geometry while the simulation assumes only right-handed rotations. Embedded in the simulation is a script, \texttt{plugin\_APS\_34IDC} which converts these 34-ID-C conventions to the simulation's right-handed system for simplicity and generates the fundamental BCDI rotation matrices described in Section \ref{section:Simulation_Geometry}. The simulation can be easily adapted to any beamline.
    
    \subsection{Space Transformation Scripts} \label{subsection:Space_Transformation_Scripts}
    There are three MATLAB scripts that allow the transformation between various spaces involved in the BCDI reconstruction process. The inputs and outputs are all 3D complex MATLAB arrays. Two of the scripts have the option to create binary masks (based on a thresholded input amplitude) in DCS for a specific $hkl$ of an object. The first, \texttt{SS\_to\_DCS}, will allow a reader to create a mask from a previously constructed SS shape, and the second, \texttt{DCS\_to\_DCS}, allows creation of a mask from a previously constructed DCS shape for a different $hkl$ reflection. These phasing masks can be turned into ``supports'' by convolution with a Gaussian and thresholding to ensure that the support is about 10\% larger than the mask. A support is often used as a starting guess in phase retrieval algorithms, where the phase of the diffraction pattern is obtained by mapping between DCS and DRS, applying real and reciprocal constraints, and updating the support every few iterations. These supports are used to set the outside amplitude to zero, since the object is assumed to be contained \cite{Newton2010}. Having scripts to produce supports based on prior knowledge (e.g. a sample shape or measured reflection) is expected to aid in the convergence of solutions and produce more accurate reconstructions \cite{Marchesini2003}. The third script, \texttt{DCS\_to\_SS} transforms a DCS shape to SS. Transformations to and from the SS is particularly useful for MBCDI, which involves reconstructions from three or more different reflections to recover the full lattice strain tensor \cite{Hofmann2017b,Hofmann2018}. By setting the sample motor angles to their zero positions, one can recover the LS shape instead of the SS shape, as explained in Section \ref{subsection:Sample_Space_to_Lab_Space}. Each of these scripts uses the beamline-dependent plugin as used in the simulation. We are working on creating plugins for other beamlines — contact the authors if you would like this code to be extended to your instrument.

\section{Space Transformations} \label{section:Space_Transformations}
\begin{figure} \label{fig:spaces}
    \centering
    \includegraphics[width=\textwidth,scale=0.5]{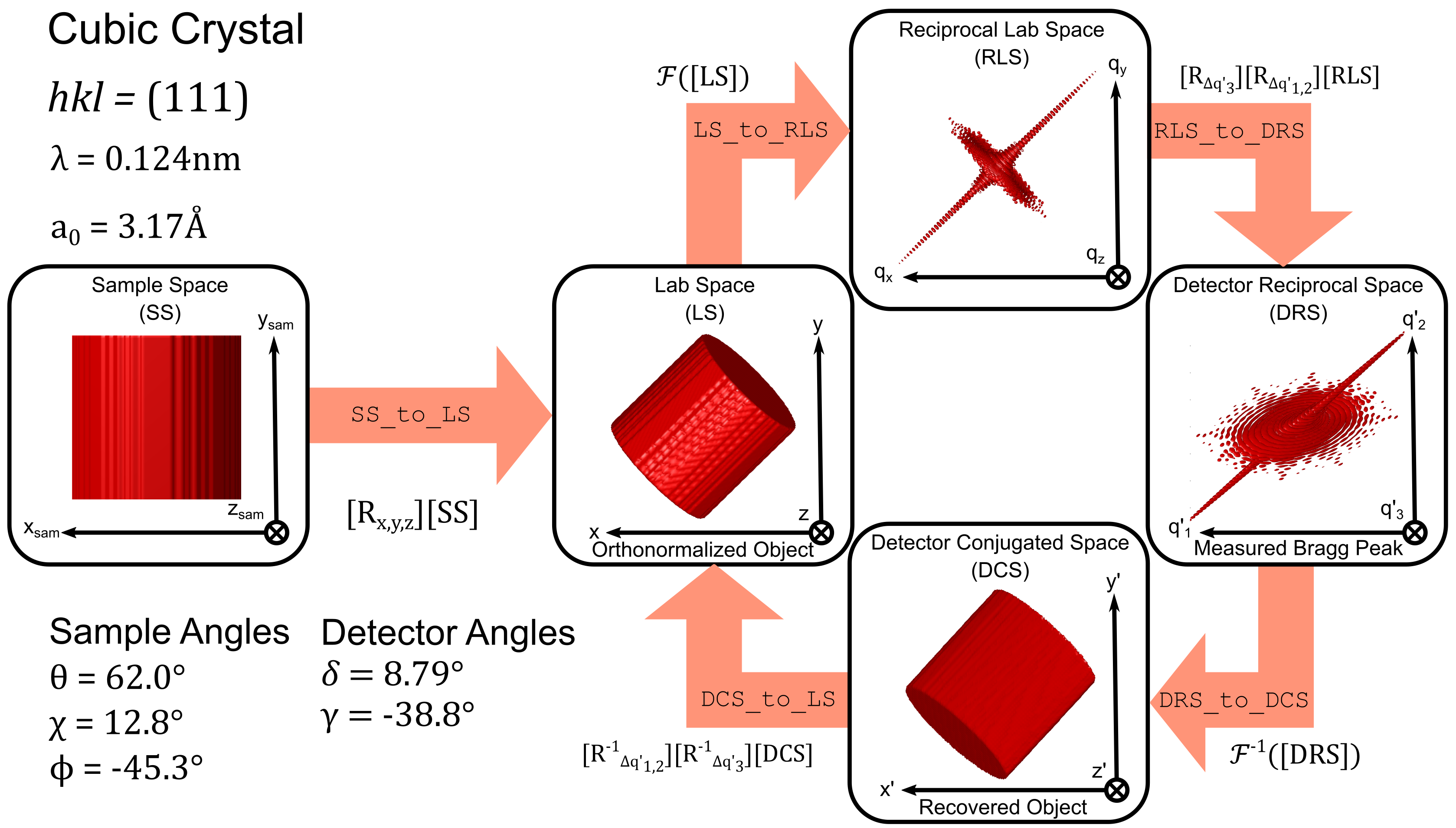} 
    \caption{Coordinate reference spaces in the MATLAB simulation for a simulated cubic crystal cylinder at $hkl = (111)$, $a_0 = 3.17\textrm{\AA}$ and $\lambda = 0.124\textrm{nm}$ based on the simulation geometry in Fig. \ref{fig:geometry}. Mathematical operations that convert from one space to another are shown, along with their respective MATLAB script names inside the arrows.}
\end{figure}

\sloppy This section pertains to the BCDI spaces in \texttt{BCDI\_{Simulation}}, shown in Fig. \ref{fig:spaces}. In the following, an outline is presented: Beginning from SS (i.e. the free standing object in its own reference frame), the goniometer rotates the specimen to LS, applying $R_\mathrm{x,y,z}$ on the SS basis vectors. The $\mathcal{F}$ is applied to LS, or equivalently, ${\mathcal{F}[\mathrm{LS}]}$, converting LS to RLS and taking the reciprocal of the LS basis vectors. The phase of the real space object is obtained as $\mathbf{\psi(r)} = \mathbf{Q}_\mathrm{lab}\cdot\mathbf{u(r)}$, which is stored in a variable rather than being recovered through an iterative algorithm. Next, the RLS is transformed to the DRS, where the measured Bragg peak lives, by applying detector and rocking matrices ${R_{\Delta \mathrm{q'_{1,2}}}}$ and ${R_{\Delta \mathrm{q'_{3}}}}$, in that order, to the RLS basis vectors. The inverse Fourier transform is applied, ${\mathcal{F}^{-1}[\mathrm{DRS}]}$, to get the DCS shape, turning the DRS basis vectors into detector conjugated real space vectors. Finally, the LS is recovered by applying the coordinate transform matrices, ${R^{-1}_{\Delta \mathrm{q'_{3}}}}$ and ${R^{-1}_{\Delta \mathrm{q'_{1,2}}}}$, in that order, to the DCS basis vectors.

To simplify, the simulation follows the following space transformation path: SS $\rightarrow$ LS $\rightarrow$ RLS $\rightarrow$ DRS $\rightarrow$ DCS $\rightarrow$ LS. For a laboratory BCDI experiment and reconstruction, the path is: DRS $\rightarrow$ DCS $\rightarrow$ LS. The SS and RLS steps are added in the simulation for clarity.

In each space the object, or its associated diffraction pattern, reside in a 3D pixel array. The basis vectors, calculated in the following subsections, capture how one moves from one pixel to the next along each axis in each space. If old basis vectors are given by $\mathbf{a}_\mathrm{1}$, $\mathbf{a}_\mathrm{2}$ and $\mathbf{a}_\mathrm{3}$ and new basis vectors are given by $\mathbf{b}_\mathrm{1}$, $\mathbf{b}_\mathrm{2}$ and $\mathbf{b}_\mathrm{3}$, the transformation matrix, $T$, can be written as

\begin{equation} \label{eq:transformation_matrix} 
        T = \begin{bmatrix} \mid\mathbf{a}_\mathrm{1}\mid  \mid\mathbf{a}_\mathrm{2}\mid  \mid\mathbf{a}_\mathrm{3}\mid \end{bmatrix} \begin{bmatrix} \mid\mathbf{b}_\mathrm{1}\mid  \mid\mathbf{b}_\mathrm{2}\mid  \mid\mathbf{b}_\mathrm{3}\mid \end{bmatrix}^{-1}.
\end{equation}

\sloppy Once the basis vectors are known, the mapping of an array from one space to another can be achieved by linear interpolation in MATLAB: \texttt{new\_array = interp3(a1grid, a2grid, a3grid, old\_array, b1grid, b2grid, b3grid)}, where \texttt{a1grid}, \texttt{a2grid} and \texttt{a3grid} are the original meshgrid coordinates with pixel number along each axis. \texttt{b1grid}, \texttt{b2grid} and \texttt{b3grid} correspond to the new meshgrid coordinates transformed by $T$ in equation (\ref{eq:transformation_matrix}). Here all basis vectors are defined in a global coordinate frame that is aligned with the lab coordinate frame. 

    \subsection{Sample Space to Lab Space} \label{subsection:Sample_Space_to_Lab_Space}
    The SS frame is the reference frame attached to the sample. This is how the sample appears on the mount, before rotating to the required geometry for any given Bragg condition. SS unit vectors are denoted by $\mathbf{\hat{x}}_\mathrm{sam}$, $\mathbf{\hat{y}}_\mathrm{sam}$, and $\mathbf{\hat{z}}_\mathrm{sam}$. To convert these unit vectors to basis vectors, we multiply by a sample pixel size, $\mathrm{p_{sam_x}}$, $\mathrm{p_{sam_y}}$ or $\mathrm{p_{sam_z}}$\footnote{As a guide, $\mathrm{p_{sam_{x,y,z}}} = \frac{\lambda D}{d N_{1,2}}$, where $\lambda$ is the X-ray wavelength, $D$ is the detector distance, $d$ is the detector pixel size and $N_1$ and $N_2$ correspond to the number of steps or pixels in the data array in LS and RLS \cite{Estandarte2018}.}

    \begin{equation} \label{eq:SS_basis}
        \begin{aligned}
            \mathbf{x}_\mathrm{sam} &= \mathrm{p_{sam_x}}\mathbf{\hat{x}}_\mathrm{sam},\\
            \mathbf{y}_\mathrm{sam} &= \mathrm{p_{sam_y}}\mathbf{\hat{y}}_\mathrm{sam},\\
            \mathbf{z}_\mathrm{sam} &= \mathrm{p_{sam_z}}\mathbf{\hat{z}}_\mathrm{sam}.\\
        \end{aligned}
    \end{equation}
    \\
    When all sample angles, $\chi$, $\phi$, and $\theta$, are zero, SS and LS are aligned. The sample is rotated by $R_\mathrm{x,y,z}$ such that it is in the LS coordinate frame. The sample rotation matrix is
    
    \begin{equation} \label{eq:R_xyz}
        R_\mathrm{x,y,z} = [R_{\mathrm{y}}(\theta)][R_{\mathrm{z}}(\chi)][R_{\mathrm{x}}(\phi)],
    \end{equation}
    
    where $R_{\mathrm{y}}$ is a right-handed rotation $\theta$ about the $y$-axis, $R_{\mathrm{z}}$ is a right-handed rotation $\chi$ about the $z$-axis, and $R_{\mathrm{x}}$ is a right-handed rotation $\phi$ about the $x$-axis. To map a position in the SS coordinate frame, $\mathbf{r}_\mathrm{SS}$, to the LS coordinate frame, i.e. find  $\mathbf{r}_\mathrm{LS}$, we simply apply equation (\ref{eq:R_xyz}) to $\mathbf{r}_\mathrm{SS}$, i.e. $\mathbf{r}_\mathrm{LS} = [R_\mathrm{x,y,z}][\mathbf{r}_\mathrm{SS}]$. Thus to obtain SS  basis vectors in the global coordinate frame, we apply
    
    \begin{equation} \label{eq:LS_basis}
        \begin{aligned}
            \relax[\mathbf{x}_\mathrm{sam}]_{\mathrm{LS}} &= [R_\mathrm{x,y,z}][\mathbf{x}_\mathrm{sam}],\\
            [\mathbf{y}_\mathrm{sam}]_{\mathrm{LS}}&= [R_\mathrm{x,y,z}][\mathbf{y}_\mathrm{sam}],\\
            [\mathbf{z}_\mathrm{sam}]_{\mathrm{LS}}&= [R_\mathrm{x,y,z}][\mathbf{z}_\mathrm{sam}].
        \end{aligned}
    \end{equation}
    \\
    The LS basis vectors, $\mathbf{x}$, $\mathbf{y}$ and $\mathbf{z}$ are aligned with the axes of the lab frame (and the global reference frame), and their size, i.e. the pixel size, should be the same as in SS because both LS and SS coordinate frames are in orthogonal real space. We use $T$ from equation (\ref{eq:transformation_matrix}) to then carry out the mapping.
    
    \subsection{Lab Space to Reciprocal Lab Space} \label{subsection:Lab_Space_to_Reciprocal_Lab_Space}
    \sloppy The LS is the reference space when a specific Bragg condition is being satisfied, in other words, as the sample sits when mounted on the goniometer at a beamline. To convert from LS to RLS, the $\mathcal{F}$ is applied to the array. For ease of simulation, prior to incorporating the position of the detector relative to the sample, we consider the complex wavefield plotted on an orthonormal reciprocal space coordinate grid. To convert from LS to RLS basis vectors, we apply the following operations,

    \begin{equation} \label{eq:LS_volume}
        \begin{aligned}
            V_\mathrm{LS} &= \mid\left(N_1 \mathbf{x} \times N_2 \mathbf{y}\right)\cdot N_3 \mathbf{z}\mid \\
            &= \mid\left(N_1 \mathrm{p_{sam_x}} \mathbf{\hat{x}} \times N_2 \mathrm{p_{sam_y}} \mathbf{\hat{y}}\right)\cdot N_3 \mathrm{p_{sam_z}} \mathbf{\hat{z}}\mid \\
            &= N_1 N_2 N_3\mathrm{p_{sam_x}p_{sam_y}p_{sam_z}} \hspace{5mm} \text{(Since $\mathbf{\hat{x}}$, $\mathbf{\hat{y}}$ and $\mathbf{\hat{z}}$ are orthogonal),}
        \end{aligned}
    \end{equation}
    
    \begin{equation} \label{eq:RLS_basis}
        \begin{split}
            \mathbf{q}_\mathrm{x} = \frac{2 \pi}{V_\mathrm{LS}} \left( N_2 \mathbf{y} \times N_3 \mathbf{z} \right) = \frac{2 \pi}{N_1\mathrm{p_{sam_x}}}\mathbf{\hat{x}},\\
            \mathbf{q}_\mathrm{y} = \frac{2 \pi}{V_\mathrm{LS}} \left( N_3 \mathbf{z} \times N_1 \mathbf{x} \right) = \frac{2 \pi}{N_2\mathrm{p_{sam_y}}}\mathbf{\hat{y}},\\
            \mathbf{q}_\mathrm{z} = \frac{2 \pi}{V_\mathrm{LS}} \left( N_1 \mathbf{x} \times N_2 \mathbf{y} \right) = \frac{2 \pi}{N_3\mathrm{p_{sam_z}}}\mathbf{\hat{z}},\\
        \end{split}
    \end{equation}
    \\
    where $V_\mathrm{LS}$ is the volume created by $\mathbf{x}, \mathbf{y}$, and $\mathbf{z}$ and $N_1$, $N_2$, $N_3$, which corresponds to the number of steps or pixels in the data array \cite{Estandarte2018}. Based on equation (\ref{eq:RLS_basis}), normalizing $\mathbf{q}_\mathrm{x}$, $\mathbf{q}_\mathrm{y}$ and $\mathbf{q}_\mathrm{z}$ would give the LS unit vectors, $\mathbf{\hat{x}}$, $\mathbf{\hat{y}}$ and $\mathbf{\hat{z}}$, respectively.
    
    \subsection{Reciprocal Lab Space to Detector Reciprocal Space} \label{subsection:Reciprocal_Lab Space_to_Detector_Reciprocal_Space}
    
    The detector will be placed at a  position corresponding to the angle required to satisfy the Bragg condition for a specific $hkl$. In DRS, diffraction patterns without phase information are collected, which will occupy non-orthonormal reciprocal space. From here, we use the $'$ symbol to represent any set of non-orthogonal basis vectors. To translate from RLS to DRS basis vectors, a coordinate transform is applied to $\mathbf{q}_\mathrm{x}$, $\mathbf{q}_\mathrm{y}$, and $\mathbf{q}_\mathrm{z}$ based on the detector position and rocking axis. For 34-ID-C, the detector is rotated by

    \begin{equation} \label{eq:R_q12}
        R_{\Delta \mathrm{q'_{1,2}}} = [R_{\mathrm{y}}(\delta)][R_{\mathrm{x}}(\gamma)],
    \end{equation}
    \\
    where $R_{\mathrm{y}}$ is a right-handed rotation $\delta$ about the $y$-axis and $R_{\mathrm{x}}$ is a right-handed rotation $\gamma$ about the $x$-axis. The rotation about the $x$-axis is performed before the rotation about the $y$-axis since the former is at the condition where $\delta=0$ \cite{Pfeifer2005}. To collect diffraction patterns, we calculate $\mathbf{q}'_\mathrm{1}$ and $\mathbf{q}'_\mathrm{2}$ that describe in-plane reciprocal space pixel positions on the detector,
    
    \begin{equation} \label{eq:q'12}
        \begin{split}
            \mathbf{q}'_\mathrm{1} = [R_{\Delta \mathrm{q'_{1,2}}}] \begin{bmatrix} \frac{\lambda D}{d N_{1}} \\ 0 \\ 0 \end{bmatrix},\\ 
            \mathbf{q}'_\mathrm{2} = [R_{\Delta \mathrm{q'_{1,2}}}] \begin{bmatrix} 0 \\ \frac{\lambda D}{d N_{2}} \\ 0 \end{bmatrix},\\
        \end{split}
    \end{equation}
    \\
    where $\lambda$ is the X-ray wavelength, $D$ is the detector distance, $d$ is the detector pixel size and $N_1$ and $N_2$ correspond to the number of detector pixels in the $x$ and $y$ directions respectively.
    
    $\mathbf{q}'_\mathrm{3}$ is determined by the rocking axis and requires the direction of the incident beam. For 34-ID-C, this is given by $\mathbf{S}_\mathrm{0 lab}$,
    
    \begin{equation} \label{eq:S_0lab}
        \mathbf{S}_\mathrm{0 lab} = \frac{2\pi}{\lambda}\begin{bmatrix} 0 \\ 0 \\ 1 \end{bmatrix}.
    \end{equation} 
    \\
    The diffracted beam, $\mathbf{S}_\mathrm{lab}$, will be incident on the detector, thus $\mathbf{S}_\mathrm{lab}$ is equal to $\mathbf{S}_\mathrm{0 lab}$ rotated by the  detector matrix $R_{\Delta \mathrm{q'_{1,2}}}$ in equation (\ref{eq:R_q12}),
    
    \begin{equation} \label{eq:S_lab}
        \mathbf{S}_\mathrm{lab} = [R_{\Delta \mathrm{q'_{1,2}}}][\mathbf{S}_\mathrm{0 lab}].
    \end{equation} 
    \\
    Finally, the scattering vector, $\mathbf{Q}_\mathrm{lab}$, is the center of the diffraction peak in reciprocal space,
    
    \begin{equation} \label{eq:Q_lab}
        \mathbf{Q}_\mathrm{lab} = \mathbf{S}_\mathrm{lab} - \mathbf{S}_\mathrm{0 lab}.
    \end{equation}
    \\
    For beamline 34-ID-C, rocking scans are performed about the $y$-axis with an increment of $\Delta \theta$ in lab coordinates. Note that we are interested in the momentum transfer in the frame of a particular crystal, so we must rotate $\mathbf{Q}_\mathrm{lab}$ by $-\Delta \theta$ \cite{Pfeifer2005}, so the rocking matrix, $R_{\Delta \mathrm{q'_{3}}}$, is
    
    \begin{equation} \label{eq:R_q3}
        R_{\Delta \mathrm{q'_{3}}} = R_{\mathrm{y}}(-\Delta \theta),
    \end{equation}
    \\
    where $R_{\mathrm{y}}$ is a right-handed rotation by $-\Delta\theta$ about the $y$-axis. Accordingly, $\mathbf{q}'_\mathrm{3}$ is
    
    \begin{equation} \label{eq:q'3}
        \mathbf{q}'_\mathrm{3} = [R_{\Delta \mathrm{q'_{3}}}][\mathbf{Q}_\mathrm{lab}] - \mathbf{Q}_\mathrm{lab}.
    \end{equation}
    \\
    Once $\mathbf{q}'_\mathrm{1}$, $\mathbf{q}'_\mathrm{2}$ and $\mathbf{q}'_\mathrm{3}$ are known, the data from the RLS can be mapped into the DRS using the transformation approach of equation (\ref{eq:transformation_matrix}).

    \subsection{Detector Reciprocal Space to Detector Conjugated Space} \label{subsection:Detector Reciprocal_Space_to_Detector_Conjugated_Space}
    \sloppy To convert from DRS to DCS, the $\mathcal{F}^{-1}$, is applied. In \texttt{BCDI\_{Simulation}}, the phase calculated when $\mathcal{F}$ is applied to the LS shape in Section \ref{subsection:Sample_Space_to_Lab_Space} is stored and used when taking the $\mathcal{F}^{-1}$. However, in a laboratory-based experiment, the phase must be recovered using a phase retrieval algorithm \cite{Clark2012} as it is lost during the detection process. Similar to DRS, DCS is a non-orthogonal space and will later need to be transformed back to LS to recover the object. The basis vectors for DCS are \cite{Berenguer2013}: 
    
    \begin{equation} \label{eq:DRS_volume}
            V_\mathrm{DRS} = \mid\left(N_1 \mathbf{q}'_\mathrm{1} \times N_2 \mathbf{q}'_\mathrm{2} \right) \cdot N_3 \mathbf{q}'_\mathrm{3}\mid, \\
    \end{equation}
        
    \begin{equation} \label{eq:DCS_basis}
        \begin{split}
            \mathbf{x}' = \frac{2 \pi}{V_\mathrm{DRS}} \left( N_2 \mathbf{q}'_\mathrm{2} \times N_3 \mathbf{q}'_\mathrm{3} \right),\\
            \mathbf{y}' = \frac{2 \pi}{V_\mathrm{DRS}} \left( N_3 \mathbf{q}'_\mathrm{3} \times N_1 \mathbf{q}'_\mathrm{1} \right),\\
            \mathbf{z}' = \frac{2 \pi}{V_\mathrm{DRS}} \left( N_1 \mathbf{q}'_\mathrm{1} \times N_2 \mathbf{q}'_\mathrm{2} \right).
        \end{split}
    \end{equation}
    \\

    \subsection{Detector Conjugated Space to Lab Space} \label{subsection:Detector_Conjugated_Space_to_Lab_Space}
    
    Using the DCS basis vectors in equation (\ref{eq:DCS_basis}) and the LS basis vectors in equation (\ref{eq:LS_basis}), a transformation matrix can be computed using equation (\ref{eq:transformation_matrix}) in order to map the object between these two spaces.

\section{Application of Mapping Scripts} \label{section:Application_Example}
\sloppy The following examples demonstrate the functionality of the scripts \texttt{SS\_to\_DCS}, \texttt{DCS\_to\_SS} and \texttt{DCS\_to\_DCS} through both the simulation of a cubic crystal cylinder and a laboratory-based strain microscopy sample composed of tungsten. 

For the simulation, the original shape in SS is a 64$\times$64$\times$64 pixel cylinder centered in a 256$\times$256$\times$256 pixel array that has been oriented for the simulated measurement of a $(111)$ reflection. This was performed at a detector distance of $1.77\textrm{m}$, a rocking increment of $0.0027\degree$ and $\lambda = 0.124\textrm{nm}$. 

The laboratory-based measurement is from an experiment at 34-ID-C for the $(\bar{1}10)$ reflection from a tungsten micro-crystal and is included to demonstrate the application of the scripts. It was performed at a detector distance of $1.75\textrm{m}$, a rocking increment of $0.005\degree$ and $\lambda = 0.124\textrm{nm}$.

    \subsection{Transfer from Detector Conjugated Space to Sample Space} \label{subsection:Transfer_DCS_to_SS}
\begin{figure} \label{fig:DCS_to_SS}
        \centering
        \includegraphics[width=\textwidth,scale=0.5]{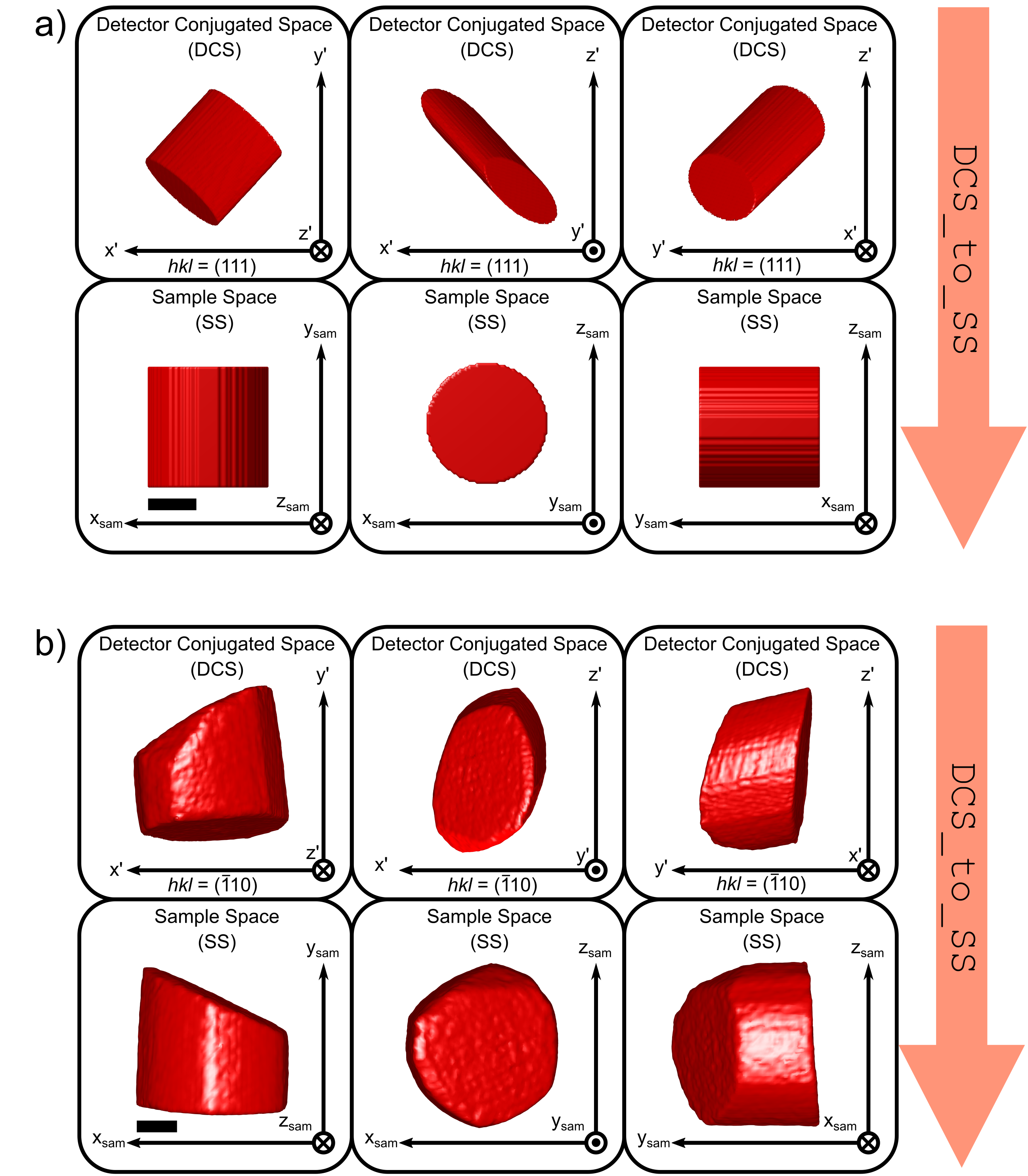} 
        \caption{Demonstration of the script \texttt{DCS\_to\_SS} for both (a) simulation and (b) laboratory-based sample, displayed for the $x$-$y$, $x$-$z$ and $y$-$z$ planes from left to right. The DCS shape is a distorted version of the SS shape. After applying \texttt{DCS\_to\_SS}, the DCS shape has been correctly converted to the initial sample in SS, also shown in Appendix \ref{appendix:a2.2} Fig. \ref{fig:lab_sample_micrographs}. The scale bars in SS correspond to $200\textrm{nm}$.}
    \end{figure}

    Fig. \ref{fig:DCS_to_SS} shows the transformation of simulated and laboratory-based DCS shapes to SS. It should be noted that the solution given by iterative phase retrieval possesses trivial non-uniqueness, where either the object or the twin of the object are valid solutions. Practically, it is normally possible to distinguish the true object from its twin by the orientation of the shape. In the case of the twin, one must take the conjugate of the $\mathcal{F}$ of the object. For the laboratory-based example in Fig. \ref{fig:DCS_to_SS}(b), the SS shape matches with the scanning electron microscope (SEM) images of the sample sitting on the mount shown in Appendix \ref{appendix:a2.1}, Fig. \ref{fig:lab_sample_micrographs}, indicating that it is indeed the correct shape and orientation. Furthermore, SS shapes calculated from two different reflections, $(\bar{1}\bar{1}0)$ and $(\bar{1}10)$, shown in Appendix \ref{appendix:a2.2}, Fig. \ref{fig:SS_overlap}, show a very high degree of overlap (99.84\%), as they should when the phase retrieval has successfully converged and the re-mapping into SS frame has been carried out correctly. 
    
    Note that the overlap, even for the simulation, is not 100\%. For the simulation, the primary source of discrepancy is interpolation, as the object is hard-edged and is thus limited by sampling resolution. For the laboratory-based experiment, sources of error include noise, motor angle accuracy and direction-dependent resolution \cite{Cherukara2018}.

    \subsection{Transfer from Sample Space to Detector Conjugated Space} \label{subsection:Transfer_SS_to_DCS}
\begin{figure} \label{fig:SS_to_DCS}
        \centering
        \includegraphics[width=\textwidth,scale=0.5]{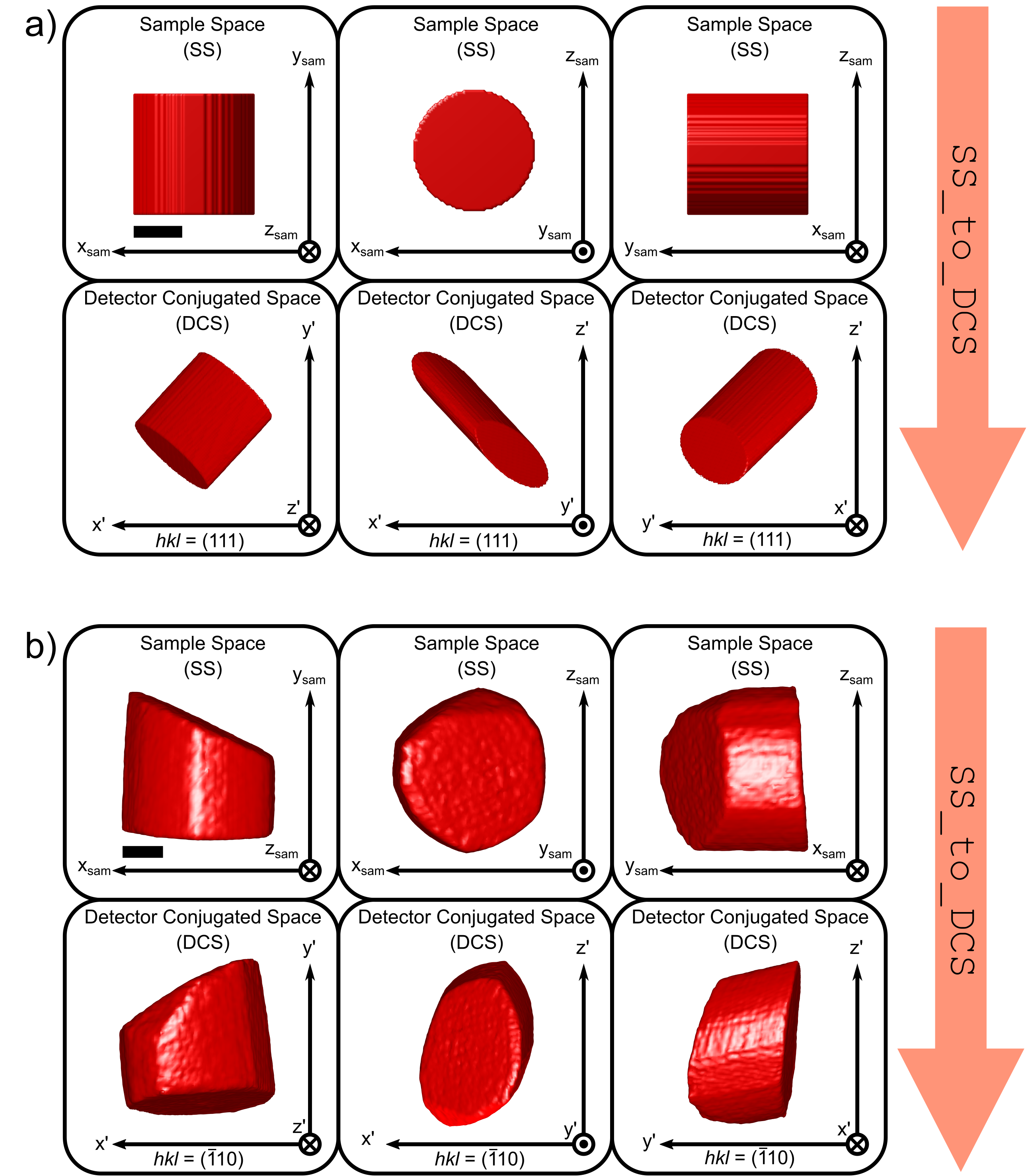} 
        \caption{Demonstration of the script \texttt{SS\_to\_DCS} for both (a) simulation and (b) laboratory-based sample, displayed for the $x$-$y$, $x$-$z$ and $y$-$z$ planes. The SS shape is distorted to become the DCS shape after applying \texttt{SS\_to\_DCS}. The scale bars in SS correspond to $200\textrm{nm}$.}
    \end{figure}

    Fig. \ref{fig:SS_to_DCS} shows the transformation of both SS shapes to DCS. The outputs of the script match the inputs of the \texttt{DCS\_to\_SS} in Fig. \ref{fig:DCS_to_SS}. For the laboratory-based example in Fig. \ref{fig:SS_to_DCS}(b), the calculated DCS shape matches the reconstructed DCS shape in Appendix \ref{appendix:a2.3}, Fig. \ref{fig:DCS_overlap}(b). The ability to map a shape from SS to DCS is important as it allows a support for reconstruction to be made, e.g. based on multiple-view SEM micrographs.
    
    \subsection{Transfer from Detector Conjugated Space to Detector Conjugated Space} \label{subsection:Transfer_DCS_to_DCS}
\begin{figure} \label{fig:DCS_to_DCS}
        \centering
        \includegraphics[width=\textwidth,scale=0.5]{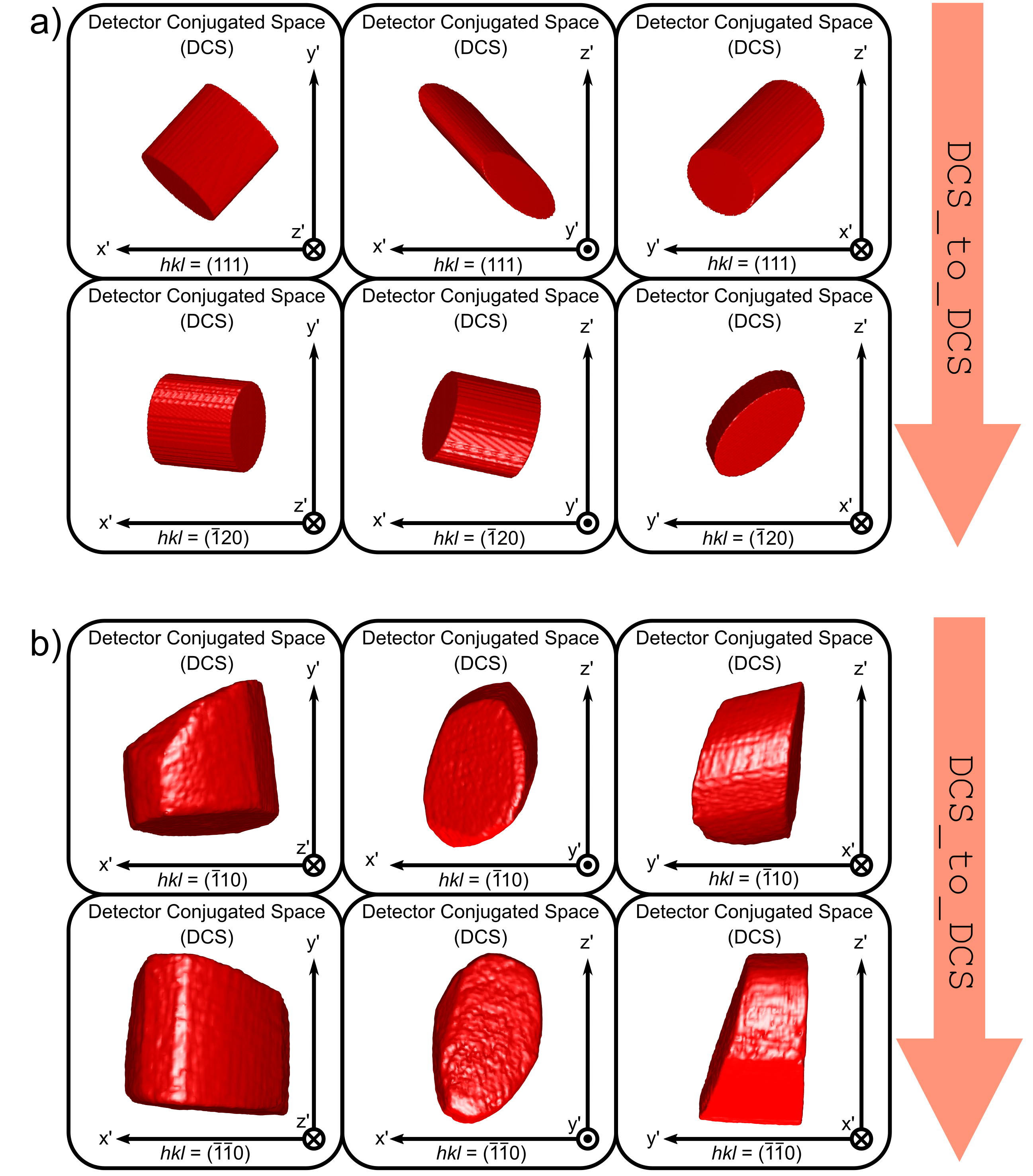} 
        \caption{Demonstration of the script \texttt{DCS\_to\_DCS} for both (a) simulation and (b) laboratory-based sample, displayed for the $x$-$y$, $x$-$z$ and $y$-$z$ planes from left to right. The DCS shape for one reflection is transformed to SS (shown in Fig. \ref{fig:DCS_to_SS}) and then transformed to the DCS shape for another reflection.}
    \end{figure}

    Fig. \ref{fig:DCS_to_DCS} shows the transformation of both DCS shapes to the DCS for another reflection. In Fig. \ref{fig:DCS_to_DCS}(a), the calculated output of the script matches the corresponding reconstructed DCS shape for the $(\bar{1}20)$ reflection in Appendix \ref{appendix:a2.3}, Fig. \ref{fig:DCS_overlap}(a). In Fig. \ref{fig:DCS_to_DCS}(b), the calculated output of the script matches the corresponding reconstructed DCS shape for the $(\bar{1}\bar{1}0)$ reflection shown in Appendix \ref{appendix:a2.3}, Fig. \ref{fig:DCS_overlap}(b).
    
\section{Conclusion} \label{section:Conclusion}
The program, \texttt{BCDI\_Simulation}, provides a flexible framework for mapping data between different coordinate frames in BCDI experiments. Importantly, it is easily adapted to the specific geometry used at different instruments. We generalize the inputs to three fundamental matrices: ${R_\mathrm{x,y,z}}$, ${R_{\Delta \mathrm{q'_{1,2}}}}$, and ${R_{\Delta \mathrm{q'_{3}}}}$, corresponding to the rotation of the sample, detector and rocking axis respectively.

As a lensless technique, BCDI's limiting factor is the maximum spatial frequency at which the phase can be reliably recovered and the durability of the algorithm to reconstruct 3D objects \cite{Xiong2014}. Scripts \texttt{SS\_to\_DCS} and \texttt{DCS\_to\_DCS} provide a means to create more accurate DCS shapes, which could be turned into initial supports for phase retrieval. This will be particularly important for samples with complex geometry and/or strong phase variations where ab initio methods begin to stagnate. The final tool, \texttt{DCS\_to\_SS}, allows the reader to map the recovered sample shape into an orthogonal sample space. This is essential for MBCDI where multiple reflections from the same object are measured and must be projected into a common coordinate frame for analysis. We expect that these tools will be useful for the community in exploring data collection in different geometries, as well as new reconstruction approaches.

\pagebreak
\appendix
\section{Appendix} \label{appendix}
    \subsection{34-ID-C Beamline to Right-Handed Conversion} \label{appendix:a1}
    
    The motor angles from beamline 34-ID-C are converted to right-handed angles in Table \ref{table:angle_conversion}, and the accessible motor ranges are noted. In Table \ref{table:angle_conversion}, $\mathrm{angle_{bl}}$ refers to the angle recorded in SPEC at 34-ID-C. 
    
    \begin{table} \label{table:angle_conversion}
            \caption{Converting from 34-ID-C (APS) $\mathrm{beamline}$ angles to right-handed angles}
            \begin{tabular}{l|l}      
            Right-handed sample motor angles & Sample motor limits \\
            \hline
             $\theta = \theta_{\mathrm{bl}}$ & $-180\degree<\theta_{\mathrm{bl}}<180\degree$\\
             $\chi = 90-\chi_{\mathrm{bl}}$ & $-15\degree<\chi_{\mathrm{bl}}<15\degree$\\
             $\phi = \phi_{\mathrm{bl}}$ & $-17\degree<\phi_{\mathrm{bl}}<17\degree$\\\\
            Right-handed detector motor angles & Setector motor limits \\
            \hline
             $\delta = \delta_{\mathrm{bl}}$ & $-1\degree<\delta_{\mathrm{bl}}<45\degree$\\
             $\gamma = 90-\gamma_{\mathrm{bl}}$ & $-1\degree<\gamma_{\mathrm{bl}}<90\degree$\\
        \end{tabular}
    \end{table}
    
    \subsection{Laboratory Example Additional Images} \label{appendix:a2}
    An MBCDI experiment was performed at 34-ID-C. The sample was a tungsten ($a_0 = 3.17\textrm{\AA}$) microcrystal prepared by a focused ion beam (FIB) liftout procedure \cite{Hofmann2019} with $\lambda = 0.124\textrm{nm}$. The recovered amplitudes for two reflections, $(\bar{1}\bar{1}0)$ and $(\bar{1}10)$, are shown in this section. The right-handed detector angles for the reflections are listed in below.
        
    \begin{table} \label{table:lab_experiment_angles}
        \caption{Laboratory-based MBCDI experiment right-handed angles}
        \begin{tabular}{l|l}      
            $(\bar{1}\bar{1}0)$ reflection &$(\bar{1}10)$ reflection \\
            \hline
             $\theta_{(\bar{1}\bar{1}0)} = 96.70\degree$ & $\theta_{(\bar{1}10)} = -174.2\degree$\\ 
             $\chi_{(\bar{1}\bar{1}0)} = -0.9938\degree$ & $\chi_{(\bar{1}10)} = 0.9590\degree$\\
             $\phi_{(\bar{1}\bar{1}0)} = 0.1112\degree$ & $\phi_{(\bar{1}10)} = 9.954\degree$\\
             $\delta_{(\bar{1}\bar{1}0)} = 32.26\degree$ & $\delta_{(\bar{1}10)} = 29.58\degree$\\
             $\gamma_{(\bar{1}\bar{1}0)} = -2.132\degree$ & $\gamma_{(\bar{1}10)} = -13.39\degree$
        \end{tabular}
    \end{table}
    
    \subsubsection{Electron Micrograph Comparison} \label{appendix:a2.1}
\begin{figure} \label{fig:lab_sample_micrographs}
        \centering
        \includegraphics[width=\textwidth,scale=0.5]{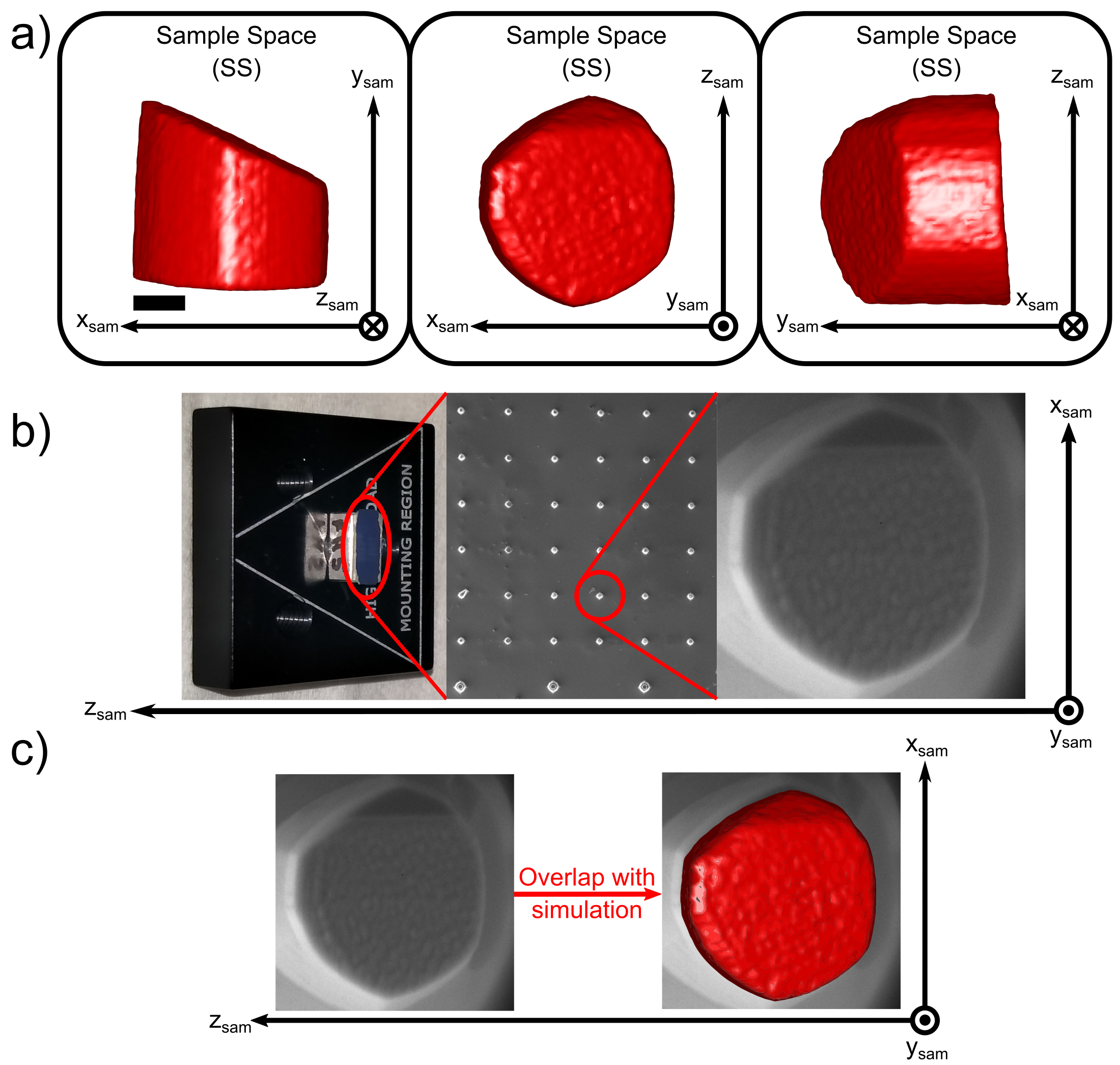} 
        \caption{Comparison between (a) the calculated SS shape from the  $(\bar{1}10)$ reflection and (b) an electron micrograph of the sample. Excellent overlap between the calculated SS shape and the SEM micrograph is shown in (c). The unique shape of this object inspires confidence in the orientation of the object and thus in the faithfulness of the rotations applied to recover the SS object. The scale bar in SS corresponds to $200\textrm{nm}$.}
    \end{figure}

    The SS shape calculated from the $(\bar{1}10)$ DCS shape using \texttt{DCS\_to\_SS}, and a comparison with an SEM image of the sample is shown in Fig. \ref{fig:lab_sample_micrographs}.
    
    \subsubsection{Laboratory SS Shape Comparison}\label{appendix:a2.2}
    
\begin{figure} \label{fig:SS_overlap}
        \centering
        \includegraphics[width=\textwidth,scale=0.5]{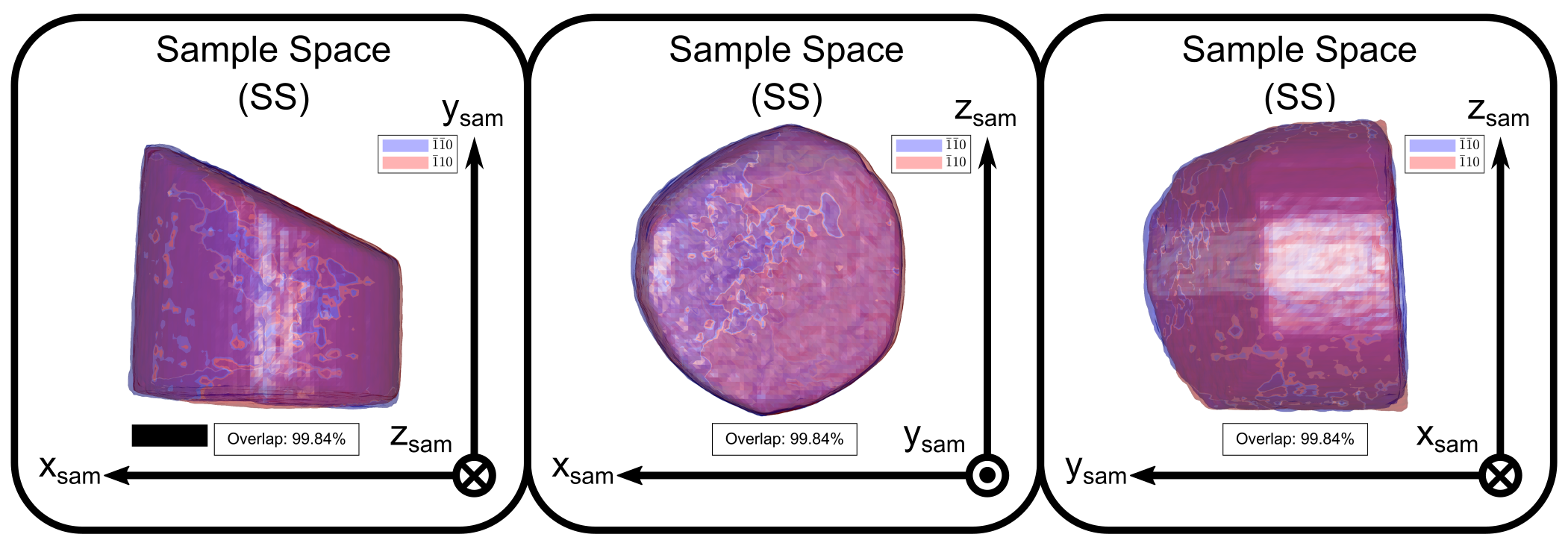} 
        \caption{Comparison between calculated SS shapes from two different reflections, $(\bar{1}\bar{1}0)$ and $(\bar{1}10)$, displayed for the $x$-$y$, $x$-$z$ and $y$-$z$ planes from left to right for a tungsten strain-microscopy sample. The 99.8\% amplitude agreement overlap highlights the accuracy of the scripts in a laboratory-based experiment. The scale bar in SS corresponds to $200\textrm{nm}$.}
    \end{figure}

    The SS shapes from two laboratory-based reflections, $(\bar{1}10)$ and $(\bar{1}\bar{1}0)$, are calculated using \texttt{DCS\_to\_SS}. The two SS objects are overlapped and viewed in the $x$-$y$, $x$-$z$ and $y$-$z$ planes in Fig. \ref{fig:SS_overlap}, showing a $99.84\%$ overlap, indicating a high accuracy of the experiment and the script, mentioned in Section \ref{subsection:Transfer_DCS_to_SS}.
    
    \subsubsection{DCS Shape Reconstruction} \label{appendix:a2.3}
\begin{figure} \label{fig:DCS_overlap}
        \centering
        \includegraphics[width=\textwidth,scale=0.5]{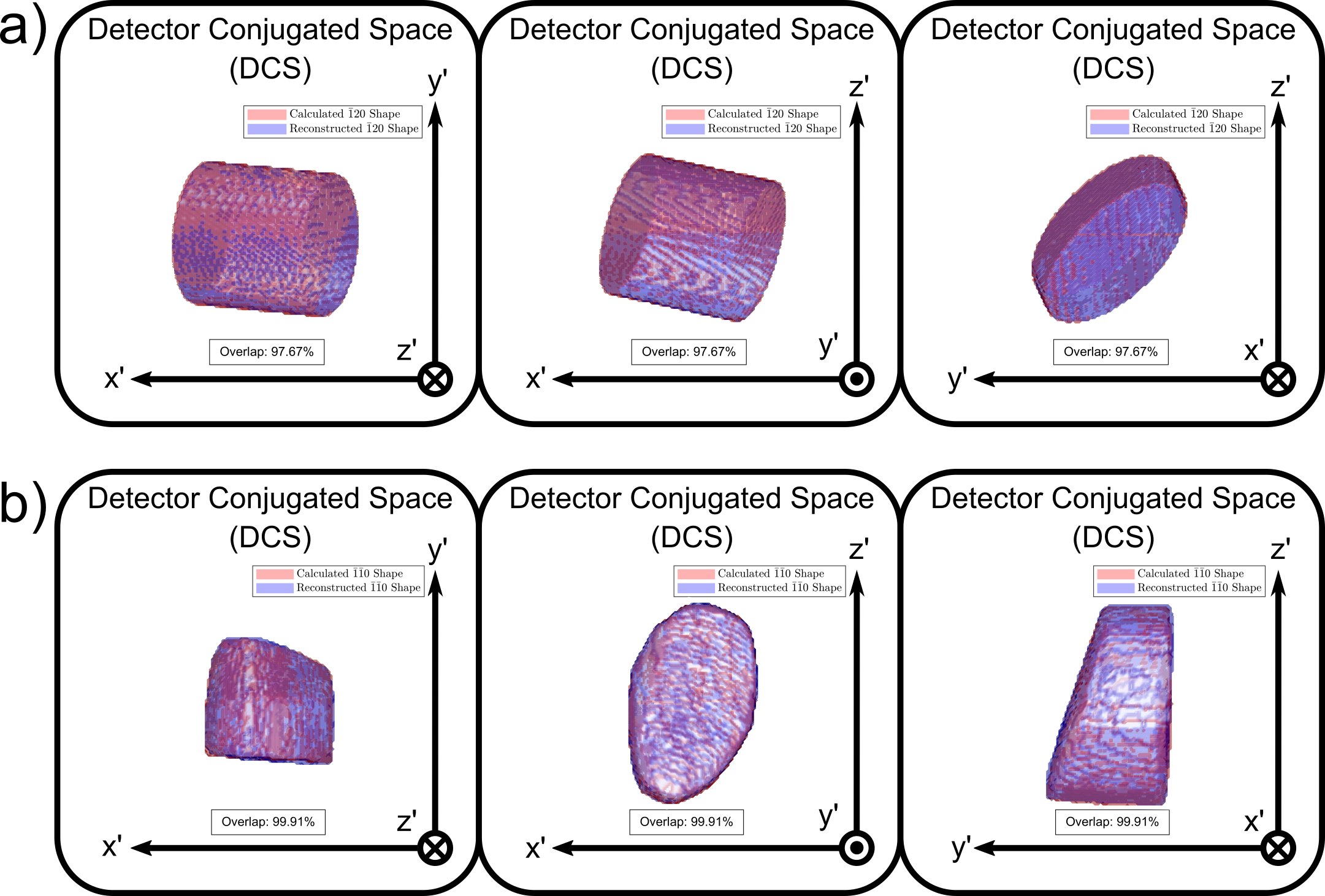} 
        \caption{Comparison between calculated DCS shape and reconstructed DCS shape for (a) simulated cylinder and (b) laboratory sample, displayed for the $x$-$y$, $x$-$z$ and $y$-$z$ planes from left to right. It shows a 97.7\%  and a 99.9\% amplitude agreement overlap respectively, highlighting the accuracy of the scripts relative to a reconstructed shape.}
    \end{figure}

    For the cylinder, the reconstructed DCS shape for $(111)$ was transformed to the DCS shape for $(\bar{1}20)$ using \texttt{DCS\_to\_DCS} and compared to the reconstructed DCS shape for $(\bar{1}20)$ in Fig. \ref{fig:DCS_overlap}(a). For the laboratory sample, the reconstructed DCS shape for $(\bar{1}10)$ was transformed to the DCS shape for $(\bar{1}\bar{1}0)$ using \texttt{DCS\_to\_DCS} and compared with the reconstructed DCS shape for $(\bar{1}\bar{1}0)$ in Fig. \ref{fig:DCS_overlap}(b). They show a $97.7\%$ and $99.9\%$ overlap respectively, indicating a high accuracy of the experiment and the script, mentioned in Section \ref{subsection:Transfer_SS_to_DCS}.
    
\ack{\textbf{Acknowledgements}} \label{acknowledgements}
\sloppy The authors would like to thank Ross Harder and Wonsuk Cha for elucidating the 34-ID-C beamline geometry, and Xiaojing Huang for clarifying the detector conjugated space to lab space mapping. This work was funded by the European Research Council under the European Union's Horizon 2020 research and innovation programme (grant agreement No 714697 to FH). Diffraction experiments were performed at the Advanced Photon Source, a US Department of Energy (DOE) Office of Science User Facility operated for the DOE Office of Science by Argonne National Laboratory under Contract No. DE-AC02-06CH11357. The authors would like to acknowledge the use of the University of Oxford Advanced Research Computing (ARC) facility in carrying out this work (http://dx.doi.org/10.5281/zenodo.22558).



\bibliographystyle{iucr}
\referencelist{iucr}

\end{document}